\documentclass[preprint,amsmath,amssymb]{revtex4}

\usepackage{graphicx}% Include figure file 
\usepackage{dcolumn}% Aligntable columns on decimal point 
\usepackage{bm}% bold math
\usepackage[latin1]{inputenc} \newcommand{\comment}[1]{}
\newcommand\etal{\mbox{\textit{et al.~}}}

\begin{document}
\setlength{\unitlength}{0.7\textwidth} \preprint{}

\title{Inertial range scaling of the scalar flux spectrum in
  two-dimensional turbulence}

\author{W.J.T. Bos$^1$, B. Kadoch$^2$, K. Schneider$^2$ and J.-P. Bertoglio$^1$}

\affiliation{$^1$ LMFA-CNRS, Universit\'e de Lyon, Ecole Centrale de Lyon,
  Universit\'e Lyon 1, INSA Lyon, 69134 Ecully, France\\
$^2$ M2P2, CNRS \& CMI, Universit\'es d'Aix-Marseille, 13453 Marseille, France
}

\begin{abstract}
Two-dimensional statistically stationary isotropic turbulence with an imposed uniform scalar gradient is investigated. Dimensional arguments are presented to predict the inertial range scaling of the turbulent scalar flux spectrum in both the inverse cascade range and the enstrophy cascade range for small and unity Schmidt numbers. The scaling predictions are checked by direct numerical simulations and good agreement is observed.  
\end{abstract}

%\pacs{47.27.Ak, 47.27.Eq, 47.27.Gs, 47.27.Jv, 47.27.Te, 47.27.Qb}
\maketitle

\section{introduction}

In the present work we consider the spectral distribution of the
passive scalar flux in two-dimensional incompressible Navier-Stokes turbulence. The scalar flux appears as the unclosed quantity in the Reynolds averaged equation for the mean scalar field: separating the velocity and passive scalar field into mean and fluctuations, $\bm u=\bar{\bm u}+\bm u'$ and $\theta=\bar \theta+\theta'$, the equation for the mean scalar field reads
\begin{equation}
\frac{\partial \bar\theta}{\partial t}+\bar u_j\frac{\partial \bar \theta}{\partial x_j}=\kappa\frac{\partial^2 \bar \theta}{\partial x_j^2}-\frac{\partial \overline{u_j'\theta'}}{\partial x_j}, 
\end{equation}
where $\kappa$ is the diffusivity of the scalar and $\overline{~\cdot~}$ denotes an ensemble average.  The last term of this equation contains the correlation
$\overline{u_j'\theta'}$, which is called the scalar flux. It is the
term which represents the influence of the turbulent fluctuations on
the mean scalar profile. Since it is the unclosed term in the Reynolds
averaged equations, it needs to be modeled, \emph{e.g.} by means of an
eddy diffusivity. To propose correct models for the scalar flux,
understanding of the physics of the turbulent flux is needed. For an
overview of models for the scalar flux, we refer to the book by Schiestel \cite{Schiestel}, the work by Rogers \cite{RogersNASA} or more recently the model derived by Wikstr\"om \etal \cite{Wikstrom2000}. For the more complicated case of the scalar flux in the presence of shear and rotation, see the work by Brethouwer \cite{Brethouwer2005}. These studies focus on three-dimensional turbulence.

We consider statistically homogeneous velocity and scalar fields so that we can investigate the scale-distribution of the turbulent scalar flux by means of Fourier spectra. The Fourier spectrum related to the scalar flux is defined as 
\begin{equation}\label{Fujt}
F_{u_j\theta}(k)=\int_{\Sigma(k)} \mathcal{F}|_{\bm x-\bm x'}[\overline{u_j'(\bm x,t)\theta'(\bm x',t)}]d\Sigma(k),
\end{equation}
in which $\Sigma(k)$ is a circular wavenumber shell with radius $k$, the
wavenumber, and $\mathcal{F}|_{\bm x-\bm x'}[.]$ denotes the Fourier transform with
respect to the separation vector $\bm x-\bm x'$. This definition is such that by construction we have 
\begin{equation}
\int_0^\infty F_{u_j\theta}(k)dk=\overline{u_j'\theta'},
\end{equation}
which illustrates that the scalar flux spectrum characterizes the contribution of different lengthscales (or wavenumbers) to the scalar flux. This spectrum is also called the scalar-velocity co-spectrum since it is defined as the real part of the scalar-velocity correlation in Fourier space. The imaginary part is called the quadrature spectrum. The quadrature spectrum does not contribute to the scalar flux in physical space and we therefore concentrate on the co-spectrum.

Academically the least complicated case to study the turbulent scalar flux is,  as proposed by Corrsin \cite{Corrsin}, isotropic turbulence on which we impose a stationary uniform mean scalar gradient $\partial \bar \theta /\partial x_1\equiv\Gamma$, arbitrarily chosen in the $x_1$-direction. In this case there exists one non-zero component of the scalar flux, aligned with the gradient. The other component is zero. We consider this case and in particular, we focus on the inertial range scaling of the scalar flux spectrum. We will in the following drop the subscripts and denote the co-spectrum by $F(k)$. We will also drop the primes and denote the fluctuations of velocity and scalar by $\bm u$ and $\theta$, respectively. Before starting the study of the scaling in two-dimensional turbulence, we briefly discuss the results obtained in the related case of three-dimensional turbulence. Lumley \cite{Lumley,Lumley2} predicted that at high Reynolds numbers the inertial range should fall off as $k^{-7/3}$. Indeed he predicted the inertial range to be given by
\begin{equation}
F(k)\sim\Gamma\epsilon^{1/3}k^{-7/3},
\end{equation}
with $\epsilon$ the dissipation of kinetic energy, or more precisely the energy flux at scale $k$. This scaling was investigated  experimentally in the atmospheric boundary layer \cite{Wyngaard} and in decaying grid turbulence at Taylor-scale Reynolds numbers up to $R_\lambda=600$ \cite{Mydlarski,Mydlarski2}. In these grid-turbulence experiments it was found that the $-7/3$ scaling was not observed at this Reynolds number. It was subsequently proposed \cite{Bos} that the inertial range exponent might be $-2$ instead of $-7/3$. However, in closure calculations it was shown that the $-2$ scaling was a low-Reynolds number effect and that the $-7/3$ scaling should be observed at higher Reynolds numbers \cite{Bos2005,Bos2007-1}. This was confirmed by the work of O'Gorman and Pullin \cite{GormanJFM} and recent direct numerical simulations (DNS) \cite{Watanabe2007}.

In the case of two-dimensional turbulence only few studies address the problem of the scaling of the scalar flux spectrum. Let us recall that in two-dimensional turbulence, in which the energy is injected at a wavenumber $k_i$, two cascades can be observed, first an energy cascade towards the large scales and, secondly, an enstrophy cascade to the small scales. If the injection scale is much smaller than the domain-size and much larger than the range in which the viscous stresses become important, both cascades are characterized by power-law scaling \cite{Batchelor1969,Kraichnan67,Leith4}. We focus on these inertial ranges, which we will denote by IC for the inverse energy cascade and FC for the forward enstrophy cascade range. In particular we investigate the wavenumber dependence of the scalar flux spectrum in these ranges. 

One of the few works investigating the scaling of the scalar flux spectrum in two-dimensional turbulence is \cite{Smith2002}, which mentions that the scalar flux spectrum can be roughly estimated by 
\begin{equation}\label{wrong}
F(k)\approx E(k)^{1/2}E_\theta(k)^{1/2},
\end{equation}
in which the scalar variance spectrum is defined as
\begin{equation}\label{EqEt}
E_\theta(k,t)=\frac{1}{2}\int_{\Sigma(k)} \mathcal{F}|_{\bm x-\bm x'}[\overline{\theta(\bm x,t)\theta(\bm x',t)}]d\Sigma(k).
\end{equation}
In the inverse cascade range, where both the energy spectrum $E(k)$ and the scalar variance spectrum $E_\theta(k)$ are known to obey Kolmogorov-Obukhov scaling \cite{Kolmogorov,Obukhov}, this would lead to a $k^{-5/3}$ inertial range. Close observation of the numerical results in reference \cite{Smith2002} shows that this is not the case.

In the present paper we show that this $k^{-5/3}$ inertial range prediction does not correspond to the physics of the problem. Phenomenological scalings for the inertial ranges in both the inverse cascade and the forward enstrophy cascade will be proposed for the scalar flux spectrum $F(k)$ and the scalar variance spectrum $E_\theta(k)$ for the  cases of unity and small Schmidt number 
%{\bf
(the Schmidt number is defined as the ratio of the diffusivity of momentum to that of the scalar, $Sc=\nu/\kappa$, and is identical to the Prandtl number when the passive scalar is temperature). Direct numerical simulations are carried out to verify the validity of the predictions.

%{\bf 
Note that since the scalar fluctuations are produced by a mean gradient, the scalar fluctuations are in principle not isotropic, but axisymmetric around the direction of the gradient. It was shown \cite{Herr,Gorman} that in the case of three-dimensional isotropic turbulence the spectral distribution of scalar flux can be described by a single scalar function. The distribution of scalar variance can be described by two scalar functions. In the present work, by integrating over wavenumber shells [equations (\ref{Fujt}), (\ref{EqEt})], we eliminate the angle dependence. A detailed study of the anisotropy of the scalar field will not be performed in the present work.

\section{Link between the Lagrangian timescale and scalar flux spectrum}

\begin{figure}
\setlength{\unitlength}{2cm} 
\includegraphics[width=5\unitlength]{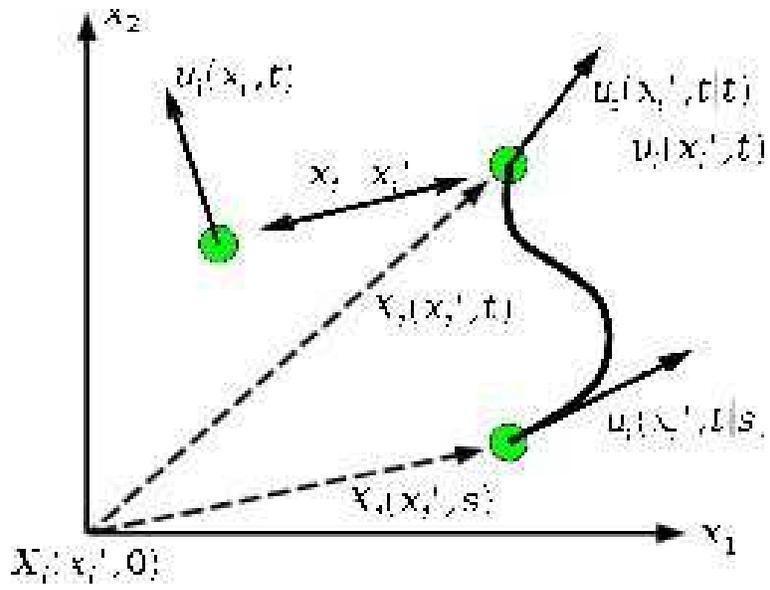}
\caption{Schematic representation of the Lagrangian two-point  velocity correlation. \label{figALHDIA}}
\end{figure}

The phenomenological scaling for the scalar flux proposed in the present work is based on the direct relation which exists between the scalar field and the Lagrangian dynamics of the turbulent velocity field. We therefore first discuss this link. Kraichnan proposed in the framework of the Lagrangian History Direct Interaction Approximation \cite{Kraichnan65}, that the dominant spectral timescale characterizing the inertial range dynamics can be estimated by
\begin{equation}\label{LHDIA}
\tau(k,t)=\int_0^{t} \frac{E(k,t|s)}{E(k,t)}ds=\frac{1}{E(k,t)}\int_0^{t} E(k,t|s)ds.
\end{equation}
This quantity was investigated numerically in \cite {Lee}. The energy spectrum is the spherically averaged Fourier transform of the two-point velocity correlation,
\begin{equation}
E(k,t)=\frac{1}{2}\int_{\Sigma(k)} \mathcal{F}|_{\bm x-\bm x'}[\overline{u_i(\bm x,t)u_i(\bm x',t)}]d\Sigma(k).
\end{equation}
$E(k,t|s)$ is the equivalent spectrum in which the Eulerian velocity $u_i(\bm x',t)$ is replaced by $u_i(\bm x',t|s)$, which is defined as the velocity at time $s$ of a fluid particle which arrives at point $\bm x'$ at time $t$. The definition of $u_i(x',t|s)$ is illustrated in Figure \ref{figALHDIA}. The definition of $E(k,t|s)$ is thus 
\begin{equation}
E(k,t|s)=\frac{1}{2}\int_{\Sigma(k)} \mathcal{F}|_{\bm x-\bm x'}[\overline{u_i(\bm x,t)u_i(\bm x',t|s)}]d\Sigma(k).
\end{equation}
By definition $E(k,t|t)$ coincides with the Eulerian spectrum $E(k,t)$. An interesting property of (\ref{LHDIA}) is that the integral can be explicited by integrating $u_i(x,t|s)$ along its trajectory. 
\begin{eqnarray}
\int_0^{t}E(k,t|s)ds=\nonumber\\ 
=\frac{1}{2}\int_{\Sigma(k)} \mathcal{F}|_{\bm x-\bm x'}[\overline{u_i(\bm x,t)\int_0^{t}u_i(\bm x',t|s)ds}]d\Sigma(k)\nonumber\\
=\frac{1}{2}\int_{\Sigma(k)} \mathcal{F}|_{\bm x-\bm x'}[\overline{u_i(\bm x,t)X_i(\bm x',t)}]d\Sigma(k).\label{Xu}
\end{eqnarray}
Instead of the two-time quantity $u_i(\bm x',t|s)$, the expression now contains the single-time displacement vector of the fluid particle, $X_i(\bm x',t)$, corresponding to the vector pointing from its position at $t=0$ to its position at $t$, $\bm x'$, or, in other words, the trajectory. The link between the scalar flux spectrum and the integral of $E(k,t|s)$ becomes evident if we compare the evolution equation of a non-diffusive passive scalar fluctuation $\theta$ in the presence of a mean scalar gradient, 
\begin{equation}\label{eqanalogy1}
\frac{\partial \theta}{\partial t}+u_j\frac{\partial \theta}{\partial x_j}= - \Gamma u_1,
\end{equation}
with the equation of the $x_1$-component of the Lagrangian position vector $X_i(\bm x,t)$:
\begin{equation}\label{eqanalogy2}
\frac{d X_1}{d t}=\frac{\partial X_1}{\partial t}+u_j\frac{\partial X_1}{\partial x_j}= u_1.
\end{equation}
Indeed, both equations are identical, only differing by a factor $-\Gamma$. As was already stated in \cite{BatchelorEuler}, the scalar fluctuation is therefore proportional to the displacement of a fluid particle in the direction of the gradient. In the limit of vanishing diffusivity, relation (\ref{LHDIA}) can thus be recasted, using (\ref{Fujt}) and (\ref{Xu}) as in \cite{Bos2006}:
\begin{equation}\label{tau2}
\tau(k)= \frac{\Gamma^{-1}F(k)}{E(k)}.
\end{equation} 
If the energy spectrum and the Lagrangian timescale are known, the
scalar flux spectrum is given by relation (\ref{tau2}). 

\subsection{Prediction of the scaling of the scalar flux spectrum at large and unity Schmidt number}

Dimensional analysis and phenomenological reasoning \cite{Kraichnan65,Tennekes} give that at a scale $l\sim k^{-1}$ the Lagrangian timescale should be approximately given by $l/u(l)$ in which the typical velocity $u(l)$ can be estimated to be of order $\sqrt{k~E(k)}$. This yields an estimation for the timescale $\tau(k)$,
\begin{equation}\label{estimtau}
\tau(k)\sim \left(k^{3}E(k)\right)^{-1/2}.
\end{equation}
Combining this relation with (\ref{tau2}) yields an estimation for the scalar flux inertial range scaling,
\begin{equation}\label{estimate}
F(k)\sim\Gamma\sqrt{\frac{E(k)}{k^3}},
\end{equation}
which is a direct relation between inertial range scaling of the scalar flux spectrum and the energy spectrum. In three-dimensional turbulence, using Kolmogorov scaling for the energy spectrum,
\begin{equation}\label{K41}
E(k)\sim \epsilon^{2/3}k^{-5/3},
\end{equation} 
leads to classical scaling for the scalar flux-spectrum,
\begin{equation}\label{Lumley}
F(k)\sim \Gamma\epsilon^{1/3}k^{-7/3}.
\end{equation}
In two-dimensional turbulence this scaling should hold in the inverse cascade range where Kolmogorov scaling is expected. In the forward enstrophy cascade range, the energy spectrum is predicted to scale as \cite{Batchelor1969,Kraichnan67,Leith4},
\begin{equation}\label{BKL}
E(k)\sim \beta^{2/3}k^{-3},
\end{equation}
with $\beta$ the flux of enstrophy in the direct cascade. This scaling was later refined introducing logarithmic corrections \cite{Kraichnan1971,Leith1972},
\begin{equation}\label{BKL2}
E(k)\sim \beta^{2/3}k^{-3}/\ln(k/k_i)^{1/3},
\end{equation}
with $k_i$ the wavenumber corresponding to the energy injection. We neglect this correction as a first approach. For this forward entrophy cascade range (\ref{estimate}) yields the scaling 
\begin{equation}\label{F2D}
F(k)\sim \Gamma\beta^{1/3}k^{-3}.
\end{equation}
It should be noted that the preceding analysis supposes a high Schmidt number. Indeed, the analogy between the position of a fluid particle and a scalar fluctuation [equations (\ref{eqanalogy1}) and (\ref{eqanalogy2})]  is exact for infinite Schmidt number. However, the effect of the Schmidt number for $Sc$ larger than one is small \cite{Herr,Zhou}. O'Gorman and Pullin \cite{GormanJFM} showed that when changing the Schmidt number from $1$ to $10^4$, the shape of the scalar flux spectrum was only little affected. We now explain this.

The equation for the co-spectrum can be derived directly from the scalar advection-diffusion equation combined with the Navier-Stokes equation (\emph{e.g.} \cite{Herr,Bos2005}). It reads
\begin{equation}\label{eqEvol}
\left[\frac{\partial }{\partial t}+(\nu+\kappa) k^2\right]F(k)=-\frac{2}{3}\Gamma E(k)+T_{u\theta}^{NL}(k).
\end{equation}
The left hand side contains the time-derivative and the influence of viscosity $\nu$ and scalar diffusivity $\kappa$. We consider the statistically stationary state in which the time-derivative term drops. The first term on the right hand side is the production of scalar flux by interaction of the velocity field with the mean scalar gradient $\Gamma$. The last term is the nonlinear interaction which contains two contributions: a purely conservative nonlinear interaction which sums to zero by integration over wavenumbers and a purely destructive pressure scrambling term which annihilates the correlation between scalar and velocity fluctuations. The viscous-diffusive term can be written as
\begin{equation}
(\nu+\kappa) k^2 F(k)=\nu(1+{Sc}^{-1})k^2 F(k).
\end{equation}
This term changes only by a factor $2$ when the Schmidt number goes from $1$ to $\infty$. The influence of the Schmidt number for $Sc$ larger than one is therefore small.

\subsection{Prediction of the scaling of the scalar flux spectrum at small Schmidt number}

In the case of  $Sc\rightarrow 0$ we do expect the above reasoning to change. We now discuss this case of small Schmidt number.

When the diffusivity becomes very large (keeping $\nu$ constant to retain an inertial range for the energy spectrum), the influence of the nonlinear terms in equation (\ref{eqEvol}) will become small, since the diffusive timescale becomes smaller than the nonlinear timescale (such as the eddy turnover time). The production term is then directly balanced by the diffusive term. In this case (\ref{eqEvol}) reduces to the equilibrium 
\begin{equation}\label{eqBalance5}
\kappa k^2 F(k)=-\frac{2}{3}\Gamma E(k),
\end{equation} 
which yields
\begin{equation}\label{eqBalance6}
 F(k)=-\frac{2\Gamma E(k)}{3\kappa k^{2}}.
\end{equation} 
O'Gorman and Pullin \cite{GormanJFM} obtained the same expression in three-dimensions. In the inverse cascade (IC) this should yield a $k^{-11/3}$ scaling, in the forward enstrophy range (FC), a $k^{-5}$ scaling. 

In section \ref{SecDNS} results of direct numerical simulations of isotropic 2D turbulence with an imposed mean scalar gradient are presented to check the relations:
\begin{equation}\label{Expressions}
\Gamma^{-1}F(k)\sim\left\{\begin{array}{ll}
\epsilon^{1/3} k^{-7/3}& \textrm{IC for } Sc\ge 1,\\
\beta^{1/3}k^{-3}& \textrm{FC for } Sc\ge 1,\\
\kappa^{-1}\epsilon^{2/3}k^{-11/3}& \textrm{IC for } Sc\ll 1,\\
\kappa^{-1}\beta^{2/3}k^{-5}& \textrm{FC for } Sc\ll 1.
\end{array}\right.
\end{equation}

\section{Predictions for the spectrum of the passive scalar variance}

It is expected that the scalar variance spectrum displays Batchelor scaling \cite{Batchelor1959} in the forward enstrophy cascade as was experimentally demonstrated by \cite{Jullien2000}, 
\begin{equation}
E_\theta(k)\sim\epsilon_\theta\beta^{-1/3}k^{-1},
\end{equation}
with $\epsilon_\theta$ the (diffusive) destruction rate of passive scalar fluctuations. In the inverse cascade, Corrsin-Obukhov scaling is expected.
\begin{equation}
E_\theta(k)\sim\epsilon_\theta\epsilon^{-1/3} k^{-5/3}.
\end{equation}
The equation for the scalar variance spectrum reads
\begin{equation}\label{eqEtvol}
\left[\frac{\partial }{\partial t}+2\kappa k^2\right]E_\theta(k)=-F(k)\Gamma+T_{\theta}^{NL}(k),
\end{equation}
with $T_{\theta}^{NL}(k)$ the nonlinear transfer term. For very small Schmidt number this equation can again be linearized, yielding for
the statistically stationary state
\begin{equation}\label{eqEtlin}
E_\theta(k)=\frac{-F(k)\Gamma}{2\kappa k^2}.
\end{equation}
This gives, using (\ref{eqBalance6}),
\begin{equation}
E_\theta(k)=\frac{E(k)\Gamma^2}{3\kappa^2 k^4}.
\end{equation}
For the scalar variance, our predictions are therefore
\begin{equation}\label{Expressions2}
E_\theta(k)\sim\left\{\begin{array}{ll}
\epsilon_\theta\epsilon^{-1/3} k^{-5/3}& \textrm{IC for } Sc\ge 1,\\
\epsilon_\theta\beta^{-1/3}k^{-1}& \textrm{FC for } Sc\ge 1,\\
\Gamma^2\kappa^{-2}\epsilon^{2/3}k^{-17/3}& \textrm{IC for } Sc\ll 1,\\
\Gamma^2\kappa^{-2}\beta^{2/3}k^{-7}& \textrm{FC for } Sc\ll 1 .
\end{array}\right.
\end{equation}

\section{Numerical verification of the proposed inertial range scalings}\label{SecDNS}

\subsection{Numerical method}

Simulations are performed using a standard pseudo-spectral method \cite{Schneider2005-3}. The simulations are fully dealiased and the resolution is $1024^2$ gridpoints for a square periodic domain of size $2\pi$. The time is advanced using a second order Adams-Bashforth time-stepping scheme. 

The equations for the vorticity field and scalar field are  
\begin{eqnarray}
\frac{\partial \omega}{\partial t}+u_j\frac{\partial \omega}{\partial x_j}=(-1)^{\alpha+1}\nu_\alpha \frac{\partial^{2\alpha} \omega}{\partial x_j^{2\alpha}}+f-\gamma\frac{\partial^{-2} \omega}{\partial x_j^{-2}} \label{eqw}\\
\frac{\partial \theta}{\partial t}+u_j\frac{\partial \theta}{\partial x_j}= (-1)^{\alpha'+1}\kappa_{\alpha'} \frac{\partial^{2\alpha'} \theta}{\partial x_j^{2\alpha'}} - \Gamma u_1
\end{eqnarray}
with the vorticity $\omega=\bm e_z \cdot (\nabla\times \bm u)$, $f$ a random-phase isotropic forcing localized in a band in wavenumber-space with a time-correlation equal to the timestep. The parameters $\alpha$ and $\alpha'$ are integers equal to one in the case of Newtonian viscosity and diffusivity and equal to $8$ in the case of hyperviscosity or hyperdiffusivity. The mean gradient $\Gamma$ is in all cases taken equal to $1$ so that the scalar flux, and its spectrum, are dominantly negative.

In all cases, hyperviscosity is used to concentrate the influence of the viscous term at the highest wavenumbers. This allows to increase the extend of the inertial range, which is the main subject in the present work. Equivalently the scalar variance is removed at the largest wavenumbers by a hyper-diffusive term except in the case of small Schmidt number. Since in that case the diffusive term becomes the dominant mechanism, the scaling is directly affected by the type of diffusion, as can be seen in expressions (\ref{eqBalance5}) and (\ref{eqEtlin}). In that case we therefore use a 'normal' Laplacian diffusive term ($\alpha'=1$). In two-dimensional turbulence the energy shows a tendency to cascade to smaller wavenumbers, \emph{i.e.}, to larger scales. To avoid a pile-up of energy at the smallest wavenumber linear Rayleigh friction (the last term in eq. (\ref{eqw})) is used, with $\gamma$ equal to unity.

Two-different fully developed turbulent flows are investigated. First the inverse cascade range, in which the forcing is localized in a wavenumber shell around $k_i=210$. In this case the forward enstrophy range is reduced to less than an octave and a full decade of inverse cascade inertial range is observed in the simulations. Second the forward enstrophy range. In this case the forcing is localized around $k_i=8$, and the inverse cascade range is absent since the friction acts strongly in the region $k<k_i$. Parameters used in the simulations are summarized in table \ref{table}. Also shown are some average values of some typical turbulence quantities.

In both velocity fields two different cases are considered for the passive scalar. One at $Sc=1$, with hyperdiffusivity ($\alpha'=8$) and one at small Schmidt number and $\alpha'=1$. It is not straightforward to define Schmidt numbers for these cases. The precise definition of the Schmidt numbers is however not important for the present study, but what is important, is the location of the inertial ranges and the ranges where diffusivity becomes important. These ranges can be determined as follows. We define a wavenumber $k_*$ at which the nonlinear timescale $\tau(k)$ becomes of the order of the diffusive timescale $(\kappa k^2)^{-1}$. If $k_*$ is in the inertial range, we can estimate its value by using expression (\ref{estimtau}) and the inertial range scalings (\ref{K41}), (\ref{BKL}). This yields $k_*\sim (\epsilon/\kappa^3)^{1/4}$ in the inverse cascade and $k_*\sim (\beta/\kappa^3)^{1/6}$ in the forward cascade. The wavenumber $k_*$ marks the crossover between an inertial-convective range and an inertial-diffusive range. We will call unity Schmidt number cases, these cases in which both viscosity and diffusivity mainly act in the last two octaves of the energy and scalar spectra, \emph{i.e}. $k_*$ is of the order of the viscous wavenumber, $(\epsilon/\nu^3)^{1/4}$.   The direct influence of the viscosity and diffusivity is then small for wavenumbers smaller than approximately $100$. In the case of small $Sc$, a normal diffusive term is used since the scaling depends directly on the Laplacian. The diffusivity is here taken large enough for it to act at all scales, including the large scales, \emph{i.e}. $k_*$ is of the order of, or smaller than $k_e$, the wavenumber at which the energy spectrum peaks.

Simulations are performed until a statistically stationary velocity field is obtained. The spectra are subsequently obtained by averaging over a time-interval of approximately $300$ time-units, until a relatively smooth spectrum is obtained. This corresponds to 270 $T_e$ for the IC-range and 540 $T_e$ for the FC-range. The large-scale turnover-time $T_e$ is here defined as $T_e=1/(k_e \sqrt{\overline{u'^2}})$.

\subsection{Results}

\begin{figure}
\setlength{\unitlength}{0.6\textwidth}
\includegraphics[width=1\unitlength]{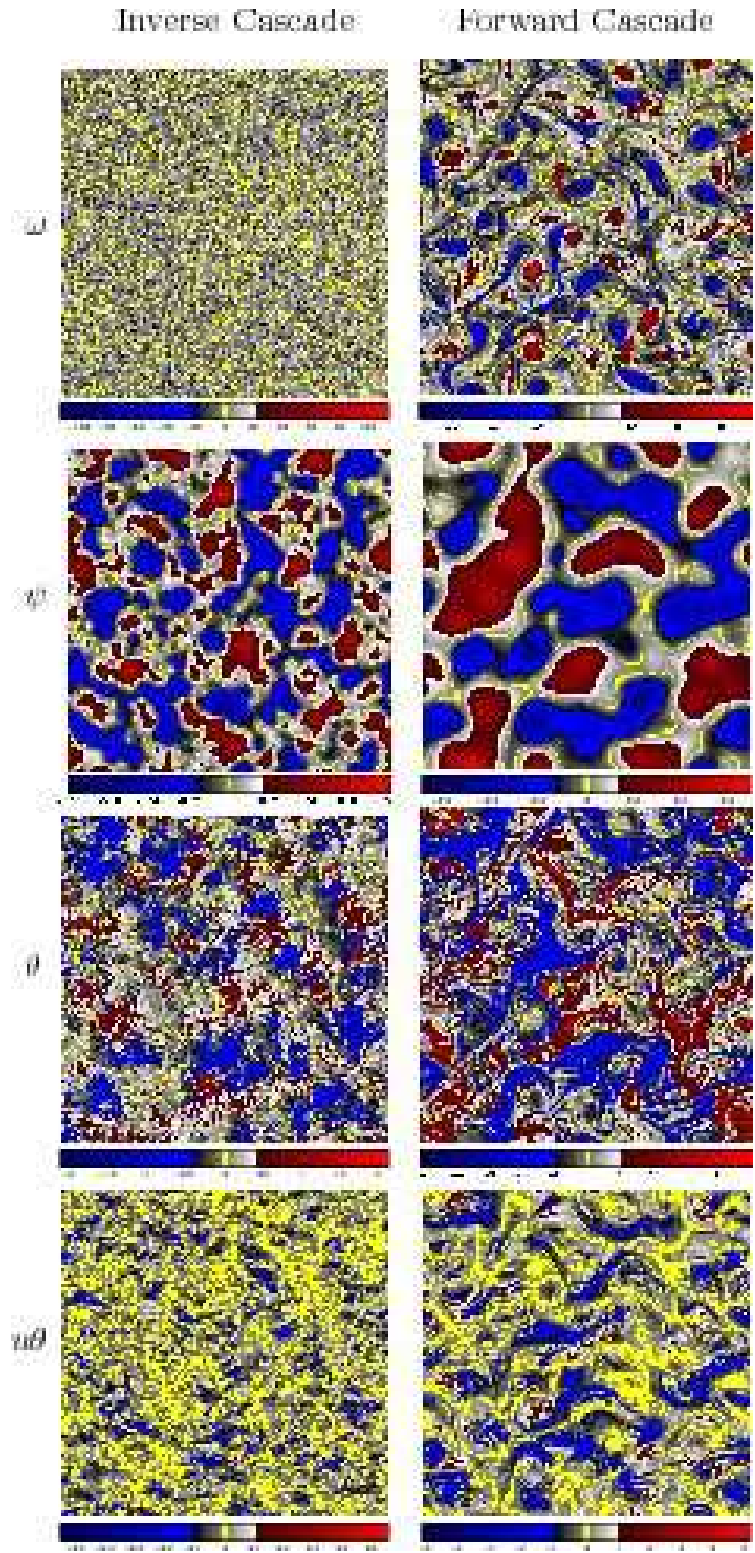}~
\caption{(Color online) Visualizations of (from top to bottom) vorticity, streamfunction, scalar fluctuations, scalar flux. Left: inverse cascade. Right: forward cascade. The Schmidt number is unity. The mean scalar gradient is in the horizontal direction.  \label{visu}}
\end{figure}

In Figure \ref{visu} visualizations of various quantities are shown at an arbitrary time. It is observed that the vorticity field contains clear vortical structures in the forward cascade. In the inverse cascade the vorticity field seems almost structureless. However, closer inspection shows small vortical structures. Visualization of the stream-function shows more clearly that these structures are present. The scalar field shows how fluctuations of passive scalar are created by interaction of the flow with the mean scalar gradient. In the IC case this scalar field is almost structureless, but shows patches of scalar fluctuation. We also displayed the instantaneous scalar flux, which is the product of the $x_1$-component of the velocity with the scalar field. Both positive and negative values of the flux are observed. The mean value is however smaller than zero (since the mean gradient is positive), so that the net flux is non-zero.

\begin{figure}
\setlength{\unitlength}{0.6\textwidth}
\includegraphics[width=1\unitlength]{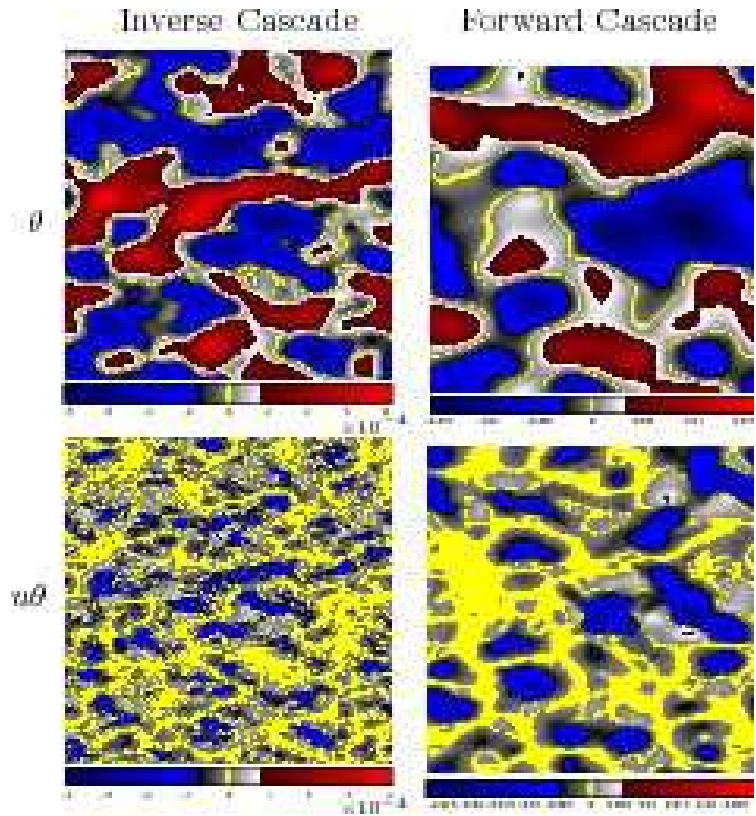}
\caption{(Color online) Visualizations of scalar fluctuations (top), scalar flux (bottom), in the inverse cascade (left) and in the forward cascade (right) for the case of small Schmidt number.\label{visu2}}
\end{figure}

In Figure \ref{visu2} visualizations are shown for the scalar field and the scalar flux for the small Schmidt number case. Vorticity fields and stream function are not shown, since they are qualitatively the same as in Figure \ref{visu}. Due to the large diffusivity, all scalar gradients are rapidly smoothed out, so that in both the IC and FC case the scalar field consists of large blobs. The scalar flux fields are characterized by a finer structure.

\begin{figure}
\begin{center}
\setlength{\unitlength}{.5\textwidth}
\includegraphics[width=1\unitlength]{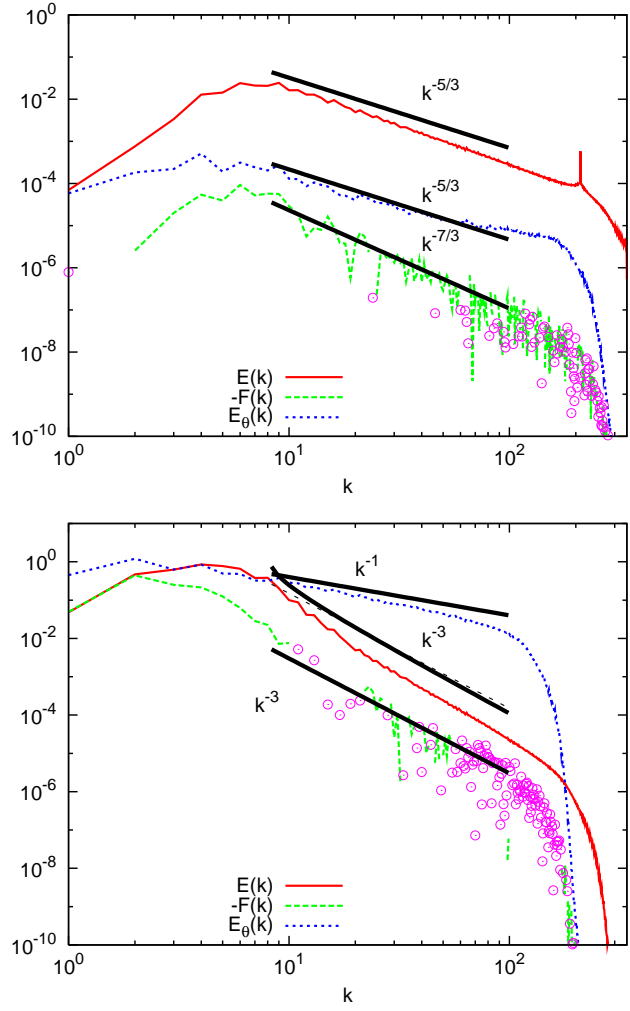}
\caption{(Color online) The energy spectrum, scalar flux spectrum and scalar variance spectrum for $Sc=1$. Top: the case of large wavenumber forcing (inverse energy cascade). Bottom: the case of small wavenumber forcing (forward entrophy cascade). The solid lines are dimensional predictions given by equation (\ref{Expressions}) and (\ref{Expressions2}). In the FC case also the log-corrected $k^{-3}$ scaling is shown for the energy spectrum, which almost superposes the normal $k^{-3}$ scaling. Dots indicate positive values of the scalar flux spectrum.
\label{Specs}}
\end{center}
\end{figure}

In Figure \ref{Specs} wavenumber spectra are shown for the energy, scalar variance and scalar flux. In the IC case, classical Kolmogorov scaling proportional to $k^{-5/3}$ holds for $E(k)$ in the inertial range. The scalar variance spectrum $E_\theta(k)$ is also proportional to $k^{-5/3}$ as can be expected from Corrsin-Obukhov arguments, but showing an important pre-diffusive bump. This bump is frequently observed in spectra of the scalar variance, \emph{e.g.} \cite{Herring,Mydlarski}. The scalar flux spectrum is proportional to $k^{-7/3}$ which is in disagreement with expression (\ref{wrong}) proposed as a rough estimate by Smith \etal \cite{Smith2002}, and in perfect agreement with expression (\ref{Expressions}), which corresponds to classical Lumley scaling. Zero-crossings are observed so that not the whole spectrum has the same sign.

In the FC range, the energy spectrum is approximately proportional to
$k^{-3}$, but slightly steeper for the wavenumbers close to the injection scale $k_i$. Taking into account the logarithmic correction, the agreement with the prediction improves even more. The scalar variance spectrum $E_\theta(k)$  shows a Batchelor regime \cite{Batchelor1959} proportional to $k^{-1}$. The scalar flux spectrum does show a scaling close to the scaling of the energy spectrum, especially for the absolute value of the spectrum. It is observed that the spectrum changes sign at several wavenumbers. These sign-changes were also observed in the investigation of the scalar flux by the stretched spiral vortex model for three-dimensional turbulence \cite{Gorman}. The spectrum of the planar contribution of the Lundgren vortex to the scalar flux showed equivalent negative excursions. We therefore relate this behavior to the roll-up of the scalar field by large coherent vortices. Indeed, a fluid particle which remains for a long time trapped in a vortical structure will contribute both positively and negatively to the scalar flux.

\begin{figure}
\begin{center}
\setlength{\unitlength}{.5\textwidth}
\includegraphics[width=1\unitlength]{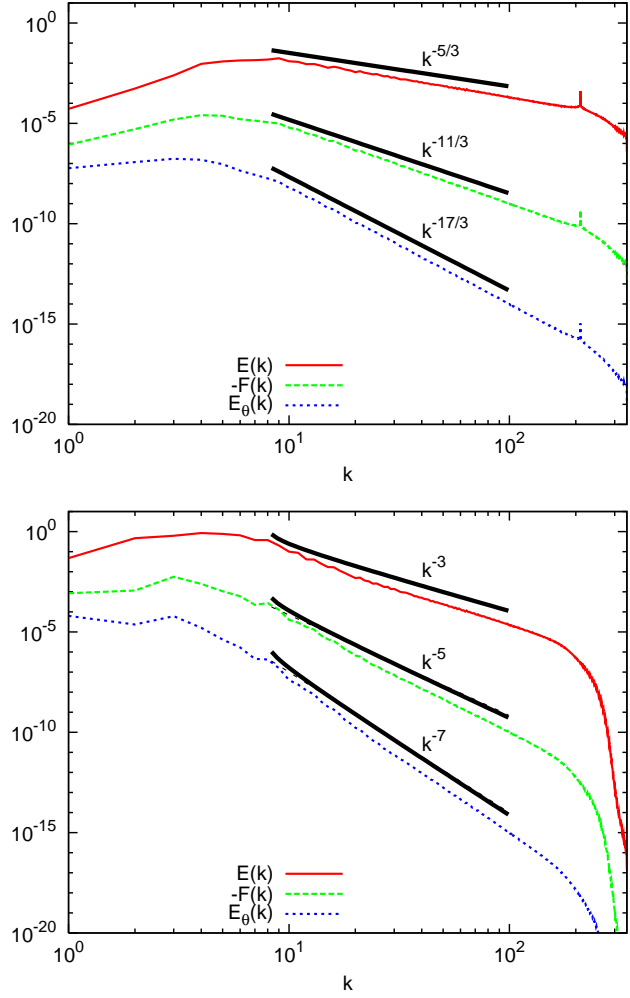}
\caption{(Color online) The energy spectrum, scalar flux spectrum and scalar variance spectrum for $Sc\ll 1$. Top: the case of large wavenumber forcing (inverse energy cascade). Bottom: the case of small wavenumber forcing (direct entrophy cascade). The solid lines are dimensional predictions given by equation (\ref{Expressions}) and (\ref{Expressions2}). In the FC case all predictions are also shown with logarithmic corrections, which almost superpose on the uncorrected scalings. \label{Specs_Pr}}
\end{center}
\end{figure}

As can be observed in Figure \ref{Specs_Pr}, at small Schmidt number, excellent agreement is observed with the predictions. In the IC range, $F(k)$ is proportional to $k^{-11/3}$ and $E_\theta(k)$ to $k^{-17/3}$. In the FC range, $F(k)$ is proportional to $k^{-5}$ and $E_\theta(k)$ to $k^{-7}$.

\section{Conclusion}

In this work the scaling of the scalar flux spectrum in two-dimensional isotropic turbulence was addressed.  Phenomenological arguments based on Lagrangian dynamics were proposed leading to the following predictions for the inertial range scaling of the scalar flux spectrum,
\begin{equation}\label{Expressions-b}
\Gamma^{-1}F(k)\sim\left\{\begin{array}{ll}
\epsilon^{1/3} k^{-7/3}& \textrm{IC for } Sc\ge 1,\\
\beta^{1/3}k^{-3}& \textrm{FC for } Sc\ge 1,\\
\kappa^{-1}\epsilon^{2/3}k^{-11/3}& \textrm{IC for } Sc\ll 1,\\
\kappa^{-1}\beta^{2/3}k^{-5}& \textrm{FC for } Sc\ll 1,
\end{array}\right.
\end{equation}
and for the scalar variance spectrum,
\begin{equation}\label{Expressions2-b}
E_\theta(k)\sim\left\{\begin{array}{ll}
\epsilon_\theta\epsilon^{-1/3} k^{-5/3}& \textrm{IC for } Sc\ge 1,\\
\epsilon_\theta\beta^{-1/3}k^{-1}& \textrm{FC for } Sc\ge 1,\\
\Gamma^2\kappa^{-2}\epsilon^{2/3}k^{-17/3}& \textrm{IC for } Sc\ll 1,\\
\Gamma^2\kappa^{-2}\beta^{2/3}k^{-7}& \textrm{FC for } Sc\ll 1 .
\end{array}\right.
\end{equation}
It was shown by DNS that in the inverse cascade the scalar flux spectrum is proportional to $k^{-7/3}$, in perfect agreement with the scaling arguments. The scalar variance shows Corrsin-Obukhov scaling, proportional to $k^{-5/3}$. In the direct enstrophy cascade the energy-spectrum obeys a log-corrected $k^{-3}$ scaling and the scalar spectrum displays Batchelor scaling proportional to $k^{-1}$. The scalar flux spectrum shows important positive and negative contributions, probably related to the presence of long-living coherent structures. The absolute value of the spectrum shows a scaling close to $k^{-3}$. At small Schmidt number, excellent agreement is observed with the predictions. The scalar flux spectrum scales here as $k^{-11/3}$ in the IC case and $k^{-5}$ in the FC case. The scalar spectrum   is proportional to $k^{-17/3}$ (IC) and $k^{-7}$ (FC).

\section*{Acknowledgments}

We thankfully acknowledge financial 
support from the Agence Nationale pour la Recherche, project 'M2TFP'.

%\bibliographystyle{report}
%\bibliography{/home/bos/PUBLI/biblio}
%\bibliographystyle{myJFM}

%\newpage

\begin{table}
\begin{tabular}{c|| c| c| c| c}
\hline
\hline
          & IC, $Sc=1$ & FC, $Sc=1$  & IC, $Sc\ll 1$ & FC, $Sc\ll 1$ \\
\hline
$k_i$ &$210$ &$8$ &$210$ &$8$\\
$k_e$ &$9$ &$4$ &$9$ &$4$\\
$\alpha$ &$8$ &$8$ &$8$ &$8$\\
$\alpha'$ &$8$ &$8$ &$1$ &$1$\\
$\nu_\alpha$ &$1.~10^{-38}$ &$1.~10^{-35}$ &$1.~10^{-38}$ &$1.~10^{-35}$\\
$\kappa_{\alpha'}$&$1.~10^{-35}$ &$1.~10^{-32}$ &$10$ &$10$\\
$\Delta t$&$5.~10^{-4}$ &$10^{-4}$ &$5.~10^{-4}$ &$10^{-4}$\\
\hline
$\overline{u^2}$&$1.~10^{-2}$ &$0.2$ &$1.~10^{-2} $ &$0.2 $\\
$\overline{v^2} $&$1.~10^{-2}$ &$0.2$ &$1.~10^{-2} $ &$0.2 $\\
$\overline{\theta^2} $&$0.1 $ &$0.9 $ &$5.~10^{-8} $ &$1.4~10^{-5} $\\
$\overline{u\theta} $&$-1.5~10^{-2} $ &$-0.1 $ &$-1.5~ 10^{-5} $ &$-1.2~10^{-3} $\\
$\overline{v\theta} $&$4.~10^{-4} $ &$2.4 ~10^{-3} $ &$-4.~10^{-8} $ &$9.~10^{-7} $\\
$\rho_{u\theta} $&$-0.45 $ &$-0.3 $ &$-0.6 $ &$-0.7 $\\
$\rho_{v\theta} $&$1.~10^{-2} $ &$6.~10^{-3} $ &$-2.~10^{-3} $ &$5.~10^{-4} $\\
\hline
\hline
\end{tabular}
\caption{\label{table}Details of the simulations. Parameters used in the simulations and average values of some typical turbulence quantities. These quantities are averaged over space and time during a time-interval of approximately $300$ time-units. The correlation coefficient $\rho_{u\theta}$ is defined as $\rho_{u\theta}=\overline{u\theta}/\sqrt{\overline{u^2}~\overline{\theta^2}}$ and analogous for $\rho_{v\theta}$.}
\end{table}

\clearpage

%============= List of Figures

\flushleft

\end{document}